\documentclass[fleqn,5p,sort&compress,times]{elsarticle}
\usepackage{graphicx}
\usepackage{hyperref}
\usepackage{color}
\usepackage{latexsym}
\usepackage{amssymb}

\newcommand{\be}{\begin{equation}}
\newcommand{\ee}{\end{equation}}
\newcommand{\ba}{\begin{eqnarray}}
\newcommand{\ea}{\end{eqnarray}}
\def\bs{\begin{subequations}}
\def\es{\end{subequations}}

\def\k{\kappa}

\def\vp{\varphi}
\def\N{\nabla}

\def\cL{\mathcal{L}}

\def\B{\Box}
\newcommand{\Eq}[1]{(\ref{#1})}
\def\com{\color{magenta}}
\def\cob{\color{blue}}
\newcommand{\book}[5]{{#1}, #2, #3, #4, #5}
\newcommand{\books}[4]{{#1}, #2, #3, #4}
\newcommand{\oarX}[1]{\href{http://arxiv.org/abs/#1}{{\ttfamily\com arXiv:#1}}}
\newcommand{\arX}[1]{\href{http://arxiv.org/abs/#1}{{\ttfamily\com arXiv:#1}}}
\newcommand{\doin}[6]{\href{http://dx.doi.org/#1}{{\cob #2 #3 {#4} (#6) #5}}}
\newcommand{\doinn}[5]{\href{http://dx.doi.org/#1}{{\cob #2 {#3} (#5) #4}}}
\newcommand{\doij}[5]{\href{http://dx.doi.org/#1}{{\cob #2 #3 (#5) #4}}}

\newcommand{\ndoinn}[5]{\href{#1}{{\cob #2 {#3} (#5) #4}}}

\newcommand{\tia}[1]{{#1},}
\def\rme{e}
\def\rmd{d}

\begin{document}

\begin{frontmatter}

\title{Stability of Schwarzschild singularity in non-local gravity}

\author{Gianluca Calcagni,}
\ead{calcagni@iem.cfmac.csic.es}
\address{Instituto de Estructura de la Materia, CSIC, Serrano 121, 28006 Madrid, Spain}

\author{Leonardo Modesto}
\ead{lmodesto@sustc.edu.cn}
\address{Department of Physics, Southern University of Science and Technology, Shenzhen 518055, China} 

\begin{abstract}
In a previous work, it was shown that all Ricci-flat spacetimes are exact solutions for a large class of non-local gravitational theories. Here we prove that, for a subclass of non-local theories, the Schwarzschild singularity is stable under linear perturbations. Thus, non-locality may be not enough to cure all the singularities of general relativity. Finally, we show that the Schwarzschild solution can be generated by the gravitational collapse of a thin shell of radiation.
\end{abstract}

\end{frontmatter}


\section{Introduction and main result}

Although there are more theories of quantum gravity than we can count with the fingers of our hands \cite{Ori09,Fousp,CQC,Car17, Modesto:2017sdr}, all share some common goals. First, to unify in one way or another the laws of quantum mechanics with a microscopic description of the gravitational force. Second, to do so in a self-consistent and predictive way, something which experts translate into the requirement of finiteness or at least renormalizability. Third, to solve some problems left open in general relativity, namely, the presence of singularities in the interior of black holes or at the birth of the universe (big bang). Fourth, to make testable predictions in the near future.

It is in this spirit, but with special attention to the third point of this agenda, that non-locality (i.e., operators with infinitely many derivatives) was proposed as a mechanism to replace the big-bang singularity with a bounce \cite{BMS}. Pursuing this hope, black-hole- or cosmology-related gravitational models with non-local operators were built \cite{BMS,ADDG,HaW,kho06,Koia,cuta8,BKM1,Bar3,Kos11,KV3,BKMV,BMMS,Kos13,Drago4,CMN,Drago3,CMT,Drago2,CMTT,EKM,Bambi:2016wdn,Drago1,KoM} together with full-fledged non-local classical and quantum gravities \cite{Kra87,kuzmin,Tom97,Bar1,Bar2,Mod11,BGKM,AMM,Mo12b,BMTs,BCKM,CaMo2,modestoLeslaw,TBM,ll16}.\footnote{There is also a whole literature on cosmological and field-theory models with non-local operators acting only on the matter sector. The reader can find more information in the papers cited above.} (For gravity in a local higher-derivative limit, see \cite{shapiro}.) In the majority of these works, the typical non-local operator, inspired by what found in string field theory, is $\sim\exp(-\Box/M^2)$, the exponential of the Laplace--Beltrami operator $\Box:=\N_\mu\N^\mu$, where $M$ is the characteristic energy scale of the theory. Indeed, many results confirmed the role of non-locality in the singularity-resolution mechanism \cite{BKM1,KV3,BKMV,Kos13,CMN,EKM}, but others were not so optimistic. For instance, singular cosmological solutions can still exist in certain general non-local theories \cite{cuta8} and conformal invariance may be an ingredient even more important than non-locality to remove infinities \cite{cuta8,Bambi:2016wdn,ll16}.

In this paper, we argue that non-locality is not enough. Despite positive results on the removal of the big-bang singularity, black-hole singularities may be more difficult to tame. There are many results about approximate singularity-free black holes in non-local gravity \cite{ModestoMoffatNico,Frolov1,Frolov2,Frolov3,Frolov4,Frolov5,Frolov6,Frolov7,Frolov8}, and the existence of the exact classical Schwarzschild black-hole solution in non-local gravitational theories was noted in \cite{yaudong}. Here we make one step further and show that this solution is stable against linear perturbations. This means that, once formed, singular spherically symmetric black holes survive just like in general relativity \cite{ReggeWheeler}.

The theory and the Schwarzschild solution are briefly reviewed in, respectively, sections \ref{sec1} and \ref{sec2}. Stability of the solution is discussed in section \ref{sec3}, while in section \ref{sec5} we show that the Schwarzschild solution can be the result of the gravitational collapse of matter. Conclusions are in section \ref{concl}.


\section{The theory}\label{sec1}

A general class of theories compatible with unitarity and super-renormalizability of finiteness has the following structure in four dimensions:
\ba
S_g = \frac{2}{\kappa^2}\int \rmd^4 x \sqrt{|g|} \left[R + R \gamma_0(\Box) R+R_{\mu\nu} \gamma_2(\Box) R^{\mu\nu} + V_g\right],\label{action}
\ea
where $\kappa^{2}=32\pi G$ and the ``potential'' term $V_g$ is at least cubic in the curvature and at least quadratic in the Ricci tensor. In this paper, we will concentrate on a reduced version of theory with $\gamma_2 =0$ in the action (\ref{action}), namely, 
\be
S_g = \frac{2}{\kappa^2}\int \rmd^4 x \sqrt{|g|} \left[R + R \, \gamma_0(\Box) R+ V_g\right].\label{action2}
\ee
These non-local theories are unitary when we choose one of the following form factors:
\ba
\gamma_0 &\hspace{-0.2cm}=\hspace{-0.2cm}& - \frac{\rme^{H(\Box)} - 1}{6\Box}\,, \label{formfactor1}\\
\gamma_0 &\hspace{-0.2cm}=\hspace{-0.2cm}& - \frac{\rme^{H(\Box)} \left(1-\frac{\Box}{M^2} \right) - 1}{6 \Box}\,, \label{formfactor2}
\ea
where $H(\Box)$ is an entire function and $M$ is a mass scale. The first form factor was first proposed by Biswas, Mazumdar and Siegel (BMS) with $H(\B)=\B$ \cite{BMS}. For convenience, we call \Eq{formfactor1} the BMS form factor for any $H$. The second form factor appears in the non-local extension of Starobinsky gravity \cite{BMTs}. The tree-level unitarity analysis for the metric perturbation around Minkowski spacetime shows that for the first choice only the spin-two graviton propagates, while for the second choice the spectrum consists in the graviton and a scalar. The theory (\ref{action2}) is non-renormal\-izable at the quantum level, however, the non-locality embodied in the form factor $\gamma_0$ implies the existence of exact regular cosmological bouncing solutions in the presence of a cosmological constant $\Lambda$ \cite{BMS,KV3,BKMV,BMMS,Kos13,Drago1,Drago2,Drago3,Drago4}, or the Starobinsky metric as an exact solution for $V_g=0$ and $\Lambda = 0$ \cite{KoshelevModestoLeslawStarob}. 


\section{Exact Ricci-flat solution}\label{sec2}

We hereby shortly recall the proof that any Ricci-flat spacetime (Schwarzschild, Kerr, and so on) is an exact solution in a large class of super-renormalizable or finite gravitational theories at least quadratic in the Ricci tensor. The equations of motion in a very compact notation \cite{Mirzabekian:1995ck} for the action (\ref{action}) read
%
\ba
E_{\mu\nu} \hspace{-0.2cm} &=& \hspace{-0.2cm} \frac{ \delta \left[  \sqrt{|g|} \left(R + R\gamma_0(\Box) R+R_{\alpha \beta} \gamma_2(\Box) R^{\alpha \beta} + { V_g}\right) \right]}{\sqrt{|g|} \delta g^{\mu\nu}} \nonumber \\
\hspace{-0.2cm} &=& \hspace{-0.2cm} G_{\mu\nu} - \frac{1}{2} g_{\mu\nu} R \gamma_0(\Box) R -  \frac{1}{2} g_{\mu\nu} R_{\alpha \beta} \gamma_2(\Box) R^{\alpha \beta}\nonumber\\
&&\hspace{-0.2cm} +2\frac{\delta R}{\delta g^{\mu \nu}} \gamma_0(\Box)R+ \frac{\delta R_{\alpha \beta}}{\delta g^{\mu \nu}} \gamma_2(\Box)R^{\alpha \beta}  \nonumber \\
&& \hspace{-0.2cm} + \frac{\delta R^{\alpha \beta}}{\delta g^{\mu \nu}  } \gamma_2(\Box)R_{\alpha \beta} +  \frac{\delta \Box^r}{\delta g^{\mu\nu} }
 \left[\frac{\gamma_0(\Box^l)-\gamma_0(\Box^r)}{\Box^r - \Box^l} R R \right]\nonumber \\
&& \hspace{-0.2cm} 	+ \frac{\delta \Box^r}{\delta g^{\mu\nu}}\left[ 
  \frac{ \gamma_2(\Box^l)- \gamma_2(\Box^r)}{\Box^r - \Box^l} R_{\alpha \beta} R^{\alpha \beta} \right]+ \frac{\delta {V_g}}{ \delta g^{\mu \nu}}\,,\label{EOM}
\ea
where $\Box^{l,r}$ act on the left and right arguments (on the right of the incremental ratio) as indicated inside the brackets.

From the above equations of motion, the following chain of implications holds in vacuum spacetime:
\ba
R_{\mu\nu} =0  &\Longrightarrow& E_{\mu\nu} = 0 \nonumber\\
&\Longrightarrow&  \mbox{Schwarzschild is an exact solution}\nonumber\\
 &\Longleftrightarrow&  \frac{\delta {V_g}}{\delta g^{\mu \nu}}= O(R_{\mu\nu})\,. 
\ea
The above result can be easily proven by replacing the \emph{Ansatz} $R_{\mu\nu} =0$ in the equations of motion (\ref{EOM}).
Of course, also the Kerr metric and all the known Ricci-flat exact solutions in vacuum of Einstein gravity are exact solutions of the non-local theory. 


\section{Stability}\label{sec3}

In this section, we study the stability of the Schwarzschild solution under linear perturbations, focusing on the theory (\ref{action2}) whose equations of motion are  obtained replacing $\gamma_2=0$ in (\ref{EOM}). Up to quadratic order in the Ricci tensor,
\be
G_{\mu\nu}+2\frac{\delta R}{\delta g^{\mu \nu}}\gamma_0(\Box) R + O({\bf Ric}^2) =0\,.\label{EOM1}
\ee
Notice that all the complicated incremental ratios in formula (\ref{EOM}) dropped out of the above equations of motion. Indeed, these ratios are quadratic in the Ricci tensor and, since we are going to consider linear perturbations in $\delta R_{\mu\nu}( h_{\alpha \beta})$, the replacement $R_{\mu\nu}(g_{\alpha \beta}) =0$ cancels them out. 
Using the explicit variation of the Ricci scalar with respect to the metric, $\delta R=(R_{\mu\nu}+g_{\mu\nu}\Box-\N_\mu\N_\nu)\,\delta g^{\mu\nu}$, the equations of motion are
\be
G_{\mu\nu} + 2(g_{\mu\nu}\Box-\N_{(\mu}\N_{\nu)})\,\gamma_0(\Box)\,R + O({\bf Ric}^2)=0\label{EOM2}\,.
\ee
Taking the trace, $R - 6 \Box \gamma_0(\Box)\, R + O(R^2) = 0$. The Einstein and trace equations for the gravitational perturbations $h_{\mu\nu}$ propagating on the Schwarzschild background read, respectively,
\ba
\hspace{-.6cm}&& \delta G_{\mu\nu} + 2(g_{\mu\nu}\Box-\N_{(\mu}\N_{\nu)})\,\gamma_0(\Box)\,\delta R=0\,,\label{EOMp}\\
\hspace{-.6cm}&& [1 - 6 \Box \gamma_0(\B)]\delta R =0\,;\label{treq}
\ea
all the other contributions are proportional to $R_{\mu\nu}$, which is zero for this specific background. 

We split the analysis in two, depending on the choice of form factor. Later on, we will unify the treatment of perturbations.


\subsection{Form factor \Eq{formfactor1}: BMS gravity}

Plugging the form factor (\ref{formfactor1}) \cite{BMS} into \Eq{treq}, we get
\be
\rme^{H(\Box)} \delta R =0\,,\label{NoSOl}
\ee
whose unique solution is\footnote{The kernel of the operator $\exp H(\Box)$ is trivial if $H$ is entire, as assumed at the beginning.} $\delta R =0$. At any point in spacetime, equation (\ref{NoSOl}) is telling us that $\delta R$ is zero with all its derivatives. Collecting together the trace equation and (\ref{EOMp}), we end up with $\delta G_{\mu\nu} =0$ and $\delta R =0$. Since $\delta G_{\mu\nu} = \delta R_{\mu\nu} - ({1}/{2}) g_{\mu\nu} \delta R$ on a Ricci-flat background, then the system is equivalent to $\delta R_{\mu\nu} =0$ and $\delta R =0$. But then again, since $R_{\mu\nu}=0$, then $\delta R =(\delta g^{\mu\nu} )R_{\mu\nu}  + g^{\mu\nu} \delta R_{\mu\nu} = g^{\mu\nu} \delta R_{\mu\nu}$, so that $\delta R_{\mu\nu} =0$ implies $\delta R=0$. Therefore, the final condition we have to solve is
\be
\delta R_{\mu\nu} =0\,.\label{EOMp4}
\ee
These are exactly the same conditions that follow from Einstein equations in vacuum and studied by Regge and Wheeler \cite{ReggeWheeler}. Therefore, one can apply those results and conclude that the Schwarzschild exact solution with line element
\ba
\rmd s^2 = g^{\rm Sch}_{\mu\nu}\rmd x^\mu\rmd x^\nu \hspace{-0.2cm} &=& \hspace{-0.2cm}  -\left(1-\frac{2Gm}{r}\right)\rmd t^2 +\left(1-\frac{2Gm}{r}\right)^{-1}\rmd r^2\nonumber\\
&& \hspace{-0.2cm} + r^2(\rmd\theta^2 + \sin^2\theta\rmd\phi^2),\label{sch}
\ea
where $m$ is the mass of the black hole, in the non-local class of theories (\ref{action2}) with form factor \Eq{formfactor1}, is stable against linear perturbations. In section \ref{gene}, we will give another simple stability argument.


\subsection{Form factor \Eq{formfactor2}: non-local Starobinsky gravity}

In this section we study again the theory (\ref{action2}), but with form factor (\ref{formfactor2}) \cite{BMTs}. The trace of the equations of motion now reads 
\be
\rme^{H(\B)} \left( 1 - \frac{\Box}{M^2} \right) \delta R = 0  \quad \Longrightarrow \quad \left( 1 - \frac{\Box}{M^2} \right) \delta R = 0\,.\label{TraceSR}
\ee
Thus, we have a massive scalar degree of freedom in the spectrum and $\delta R \neq 0$. The other components of the equations of motion are more involved than in (\ref{EOMp}), namely,
\be
\delta G_{\mu\nu} - \frac{1}{3}  \left( g_{\mu \nu} - \nabla_{\mu} \nabla_\nu \, \frac{1}{\Box} \right) \left[\rme^{H(\B)} \left( 1 - \frac{\Box}{M^2} \right) -1\right] \delta R = 0 \label{EOMpStaro}\,.
\ee
The two equations of motion (\ref{TraceSR}) and  (\ref{EOMpStaro}) must be satisfied at the same time:
\be
\left(\Box-M^2\right) \delta R = 0\,,\quad \delta G_{\mu\nu} + \frac{1}{3} g_{\mu\nu} \delta R - \frac{1}{3 M^2}  \nabla_{\mu} \nabla_\nu \delta R =0\,.\label{EOMp3Staro}
\ee
The system (\ref{EOMp3Staro}) is exactly the same one obtains from the local Starobinsky theory $\cL\propto R+R^2/(6M^2)$, in which case the stability can be proved introducing an auxiliary field at the level of the action, before perturbing \cite{Myung:2011ih}.


\subsection{General stability analysis}\label{gene}

The above stability analyses can be unified by rewriting the non-local theory (\ref{action2}) for any form factor $\gamma_0(\B)$, ignoring $V_g$, in terms of an auxiliary scalar field. Then, at a later stage one can specialize to the form factors \Eq{formfactor1} and \Eq{formfactor2}. First, we introduce a scalar field $\phi$ and afterwards we make a conformal rescaling to recast the Lagrangian in the canonical form. It is straightforward to show that (\ref{action2}) is equivalent on shell to the Lagrangian 
\ba
\hspace{-0.8cm} \mathcal{L}\hspace{-0.2cm} &=& \hspace{-0.2cm}  \frac{2}{\kappa^2} \sqrt{|g|} \left[R + \phi \frac{1}{3 m^2(\Box)} R - \phi\frac{1}{6 m^2(\Box)} \phi \right],\label{Aux1}\\
\frac{1}{m^2(\Box)} \hspace{-0.2cm} &=& \hspace{-0.2cm} 6\gamma_0(\B)\,,
\ea
i.e., when the solution of the equation of motion for the scalar field
\be
\frac{1}{m^2(\Box)}\left(R-\phi \right) = 0\,,\label{Aux1EOM}
\ee
is replaced in (\ref{Aux1}). Indeed, the most general solution of (\ref{Aux1EOM}) is
\be
\phi = R + \sum_n c_n \delta(\Box - \lambda_n)\,, 
\ee
where $\lambda_n$ are the zeros of the analytic function $1/m^2(\Box)$. However, the singular part $\sum_n  c_n \delta(\Box - \lambda_n)$ does not contribute to the action when replaced in it. Now we define 
\be
\phi =: 3 m^2 (\Box) (\rme^\vp - 1)\,, 
\ee
and make a rescaling of the metric,
\be
g_{\mu\nu} = \frac{1}{\frac{1}{3m^2(\Box)} \phi +1} \tilde{g}_{\mu\nu} = \rme^{-\vp} \tilde{g}_{\mu\nu}\,. 
\label{gtilde}
\ee
The Lagrangian for the metric $\tilde{g}_{\mu\nu}$ and the scalar $\vp$ reads 
\be
\mathcal{L} = \frac{2}{\kappa^2} \sqrt{|\tilde{g}|} \left[\tilde{R} - \frac{3}{2} (\tilde{\partial} \vp)^2 
- \frac{3}{2} \rme^{-2 \vp} (\rme^\vp -1) m^2({\Box}) (\rme^\vp -1) \right], \label{Aux2}
\ee
where $\Box = \Box(g)$ must be expressed in terms of $\tilde{g}_{\mu\nu}$ and $\vp$. 
The Schwarzschild metric $\tilde{g}_{\mu\nu} = g^{\rm Sch}_{\mu\nu}$ is an exact solution for $\vp =\vp_0\equiv0$. Therefore, we can expand the Lagrangian (\ref{Aux2}) up to second order in the fluctuation $\delta \vp$, which we can identify with $\vp$ itself because the background solution is identically zero:
\be
\vp = \vp_0 + \delta \vp = \delta \vp\,.
\ee
The action at the second order for $\vp$ reads 
\begin{eqnarray}
 \mathcal{L} \!\!\!\! & = & \!\!\!\! \frac{2}{\kappa^2} \sqrt{|\tilde{g}|} \left[\tilde{R} + \frac{3}{2} \vp \widetilde\Box \vp 
- \frac{3}{2} \vp \, m^2(\widetilde{\Box})  \, \vp  + O(\vp^3) \right]\nonumber\\
 \!\!\!\! & =: & \!\!\!\! \sqrt{|\tilde{g}|}\left[\frac{2}{\kappa^2}\tilde{R}+\frac{1}{2} \vp \frac{1}{G(\widetilde\B)} \, \vp + O(\vp^3)\right],\label{Aux2Q}
\end{eqnarray}
where $1/G(\widetilde\B)=6[\widetilde\Box-m^2(\widetilde\Box)]/\k^2$ and we approximated $\B\simeq\widetilde\B$ inside $m^2$ to leading order in $\vp$. 

Let $\widetilde\B f=-k^2 f$ be the eigenvalue equation of the d'Alembert\-ian in any given metric. The scalar-field propagator in the theory with form factor (\ref{formfactor1}),
\be
G(\widetilde\Box) = \frac{6\,\rme^{H(\widetilde\B)}-1}{\kappa^2\widetilde\Box \rme^{H(\widetilde\B)}}\,,
\ee
has a single pole in the eigenvalue $-k^2=0$ with identically zero residue. Here we used the fact that $H(0)=0$ in momentum space for all sensible form factors. Therefore, there is no scalar propagating degree of freedom on both Minkowski and Schwarzschild background and the stability of the solution is the same as in general relativity. 

For the theory with form factor (\ref{formfactor2}), the propagator reads 
\be
G(\widetilde\Box) = \frac{\rme^{H(\widetilde\B)} \left(1-\frac{\widetilde\Box}{M^2} \right) -1}{\widetilde\Box\rme^{H(\widetilde\B)} \left(1-\frac{\widetilde\Box}{M^2} \right)}.
\ee
Here there are two poles, one at $-k^2=0$ and another at $-k^2 = M^2$. The residue is again identically zero in $-k^2=0$ as long as $H(0) = 0$. The residue in $-k^2 = M^2$ is 
\be
{\rm Res} \, G(-k^2) \Big|_{-k^2 = M^2} = \frac{\rme^{- H(M^2)}}{M^2} \, . 
\ee
Therefore, the equation of motion for the scalar-field perturbation is consistent with the first equation in (\ref{EOMp3Staro}) for $\delta R$. Finally, the equation of motion for the gravitational and scalar-field perturbation around the Schwarzschild background $g^{\rm Sch}_{\mu\nu}$ and $\vp_0=0$ read
\be
\delta \tilde{G}_{\mu\nu} = 0\,, \qquad (\widetilde\Box-M^2)\vp =0\,. 
\ee
Notice that there is no contribution of the energy tensor to the Einstein's equations for the gravitational perturbations because $T_{\mu\nu}^{\vp}$ is at least quadratic in $\vp$.


\section{Gravitational collapse}\label{sec5}

In the previous section, we extensively studied the stability of the Schwarzschild metric from the mathematical point of view, assuming it is generated from the gravitational collapse of matter. In this section, we show that the Schwarzschild black hole can indeed originate from an imploding thin shell of radiation. Let us consider the Vaidya metric
\be
\rmd s^2_{\rm V} = - \left[1 - \frac{2G M(v)}{r}\right] \rmd v^2 + 2 \rmd v \rmd r +  r^2 (\rmd\theta^2 + \sin^2\theta \rmd\phi^2) \, , 
\label{Vaidya_metric}
\ee
where $M(v)$ is given by
\be
M(v) =   m\Theta (v-v_{0})\,,
\ee
where $v$ is called ingoing null coordinate. Here $m$ is the ADM mass and $\Theta$ is the unit step function (Heaviside function). For this metric, the Ricci tensor has a non-zero component but the Ricci scalar is zero, $R=0$. The energy-momentum tensor of the thin shell of radiation is
\be
T_{\mu\nu} = {\rm diag} \left( \frac{m\delta (v- v_{0})}{4 \pi  r^2}, 0, 0, 0 \right) 
\label{Tmunu} \, .
\ee
One can check that, using the metric (\ref{Vaidya_metric}), the above energy-momentum tensor is traceless, which is crucial to show that the Vaidya metric is an exact solution of the equations of motion (\ref{EOM}) for the reduced theory investigated in this paper, namely the class of theories with $\gamma_2 (\Box) =0$. In the latter case, the equations of motion simplify to 
\ba
\hspace{-.6cm}&&G_{\mu\nu} - \frac{1}{2} g_{\mu\nu} R \gamma_0(\Box) R 
 +2\frac{\delta R}{\delta g^{\mu \nu}} \gamma_0(\Box)R
  \nonumber \\
\hspace{-.6cm}&&\qquad+  \frac{\delta \Box^r}{\delta g^{\mu\nu} }
 \left[\frac{\gamma_0(\Box^l)-\gamma_0(\Box^r)}{\Box^r - \Box^l} R R \right]+ \frac{\delta {V_g}}{ \delta g^{\mu \nu}} = \frac{\kappa^{2}}{4} T_{\mu\nu}
 \,, \label{EOM3}
\ea
while their trace read
\ba
\hspace{-.6cm}&&  
-  R - 2 R \gamma_0(\Box) R 
 + 2 g^{\mu\nu} \frac{\delta R}{\delta g^{\mu \nu}} \gamma_0(\Box)R
  \nonumber \\
\hspace{-.6cm}&& \qquad
+  g^{\mu\nu} \frac{\delta \Box^r}{\delta g^{\mu\nu} }
 \left[\frac{\gamma_0(\Box^l)-\gamma_0(\Box^r)}{\Box^r - \Box^l} R R \right]+ g^{\mu\nu}\frac{\delta {V_g}}{ \delta g^{\mu \nu}} =  \frac{\kappa^{2}}{4} T = 0 \, .\nonumber\\\hspace{-.6cm} \label{EOM3trace}
\ea
The trace of the equations of motion is solved by $R=0$, provided the potential $V_g$ is at least quadratic in the Ricci scalar. Replacing $R=0$ in the full equations of motion (\ref{EOM3}), we end up with the Einstein equations sourced by a traceless energy-momentum tensor. It is well known that the Vaidya spacetime is an exact solution of the Einstein equations of general relativity; therefore, it is a solution of the non-local theory, too. 

Notwithstanding the simplicity of the dynamics studied in this section, we can conclude that the Schwarzschild metric can be the outcome of the gravitational collapse of a thin shell of radiation. Therefore, the Schwarzschild metric can be a physical solution (in the sense of coming from an astrophysical process) also in non-local theories.


\section{Conclusions}\label{concl}

In this paper, we studied the linear stability of Schwarzschild black-hole spacetime in a wide class of non-local theories, focusing then on the form factors of BMS \cite{BMS} and non-local Staro\-binsky \cite{BMTs} gravity. We found that black holes exist and are indeed linearly stable in these non-local gravities, contrary to early expectations that non-locality could resolve all the singularities of classical general relativity. It may be that different non-local theories (for instance, with Ricci-tensor non-local terms \cite{CaMo2}, $\gamma_2\neq0$ in \Eq{action}, or different form factors) avoid this result, but we do not see how at the present; the models discussed here are quite representative.

Of course, this does not exclude that a quantum version of the theories studied here could be free from black holes. However, having stable classical singular backgrounds makes this possibility remote or, at least, more difficult to realize, unless extra ingredients such as conformal invariance \cite{cuta8,Bambi:2016wdn,ll16} are allowed to enter the picture. An intriguing puzzle to solve in the future will be to understand why the same non-local theories that fail to regularize black-hole singularities are able to smoothen the big bang into a regular bounce \cite{BMS,CMN}. As the cosmological equations of motion are more complicated to analyze than Schwarzschild solutions, an answer to this problem will require more work. Moreover, in \cite{yaudong} it was also proved that the big-bang singularity persists in a large class of non-local theories for a Universe dominated by radiation. 


\section*{Acknowledgments}

G.C.\ is under a Ram\'on y Cajal contract and thanks SUSTech for the kind hospitality during the writing of this work. Both authors are supported by the I+D grant FIS2014-54800-C2-2-P.


\end{document}